\documentclass[review]{elsarticle}

\usepackage{lineno,hyperref}

\journal{Journal of \LaTeX\ Templates}

\usepackage[margin=1.7 cm]{geometry}
\usepackage{float}
\usepackage{graphicx}
\usepackage{amsmath}
\usepackage{amssymb}
\usepackage{booktabs}
\usepackage{tabularx,ragged2e}
\usepackage{xcolor}
\usepackage{caption}
\usepackage{subcaption}
\usepackage{adjustbox}
\usepackage{array}
\newcolumntype{P}[1]{>{\centering\arraybackslash}p{#1}}
\definecolor{codegreen}{rgb}{0,0.6,0}
\definecolor{codegray}{rgb}{0.5,0.5,0.5}
\definecolor{codepurple}{rgb}{0.58,0,0.82}
\definecolor{backcolour}{rgb}{0.95,0.95,0.92}

\usepackage{listings}
\lstdefinestyle{mystyle}{
	basicstyle=\footnotesize ,
    backgroundcolor=\color{backcolour},   
    commentstyle=\color{codegreen},
    keywordstyle=\color{magenta},
    numberstyle=\tiny\color{codegray},
    stringstyle=\color{codepurple},
    basicstyle=\ttfamily\footnotesize,
    breakatwhitespace=false,         
    breaklines=true,    
    postbreak=\mbox{\textcolor{red}{$\hookrightarrow$}\space},           
    captionpos=b,                    
    keepspaces=true,                 
    numbers=left,                    
    numbersep=5pt,                  
    showspaces=false,                
    showstringspaces=false,
    showtabs=false,                  
    tabsize=2
}
 
\usepackage{tikz}
 
\begin{document}

\begin{frontmatter}

\title{Space applications of GAGG:Ce scintillators: a study of afterglow emission by proton irradiation}

\author[a,b,c]{Giuseppe~Dilillo}
\author[c]{Nicola~Zampa}
\author[d,e]{Riccardo~Campana}
\author[d,e]{Fabio~Fuschino}
\author[a,c]{Giovanni~Pauletta}
\author[h]{Irina~Rashevskaya}
\author[f]{Filippo~Ambrosino}
\author[a,c]{Marco~Baruzzo}
\author[a,c]{Diego~Cauz}
\author[a,c]{Daniela~Cirrincione}
\author[a,c]{Marco~Citossi}
\author[a,c]{Giovanni~Della~Casa}
\author[h]{Benedetto~Di~Ruzza}
\author[f,g]{Yuri~Evangelista}
\author[m,n]{Gábor~Galgóczi}
\author[d,e]{Claudio~Labanti}
\author[m,l]{Jakub~Ripa}
\author[i,h]{Francesco~Tommasino}
\author[h]{Enrico~Verroi}
\author[b]{Fabrizio~Fiore}
\author[a,c]{Andrea~Vacchi}

\address[a]{University of Udine, Via delle Scienze 206, I-33100 Udine, Italy}
\address[b]{INAF-OATs Via G.B. Tiepolo, 11, I-34143 Trieste}
\address[c]{INFN sez. Trieste, Padriciano 99, I-34127 Trieste, Italy}
\address[d]{INAF-OAS Bologna, Via Gobetti 101, I-40129 Bologna, Italy}
\address[e]{INFN sez. Bologna, Viale Berti-Pichat 6/2, I-40127 Bologna, Italy}
\address[f]{INAF-IAPS,Via del Fosso del Cavaliere 100, I-00133 Rome, Italy}
\address[g]{INFN sez. Roma 2, Via della Ricerca Scientifica 1, I-00133 Rome, Italy}
\address[h]{TIFPA-INFN, Via Sommarive 14, I-38123 Trento, Italy}
\address[i]{Department of Physics, University of Trento, via Sommarive, 14 38123 Trento}
\address[l]{Department of Theoretical Physics and Astrophysics, Faculty of Science, Masaryk University, Brno, Czech Republic}
\address[m]{Eötvös Loránd University, Egyetem tér 1-3, 1053 Budapest, Hungary}
\address[n]{Hungarian Academy of Sciences, Wigner Research Centre for Physics,1525 Budapest 114, Hungary}

%



\begin{abstract}

We discuss the results of a proton irradiation campaign of a GAGG:Ce (Cerium-doped Gadolinium Aluminium Gallium Garnet) 
scintillation crystal, carried out in the framework of the HERMES-TP/SP (High Energy Rapid Modular Ensemble of Satellites --- Technological and Scientific Pathfinder) mission. A scintillator sample was irradiated with 70 MeV protons, at 
levels equivalent to those expected in equatorial and sun-synchronous low-Earth orbits over orbital periods spanning 6 months to 10 years.
The data we acquired are used to introduce an original model of GAGG:Ce afterglow emission.
Results from this model are applied to the HERMES-TP/SP scenario, aiming at an upper-bound estimate of the detector performance degradation resulting from afterglow emission.

\end{abstract}


\end{frontmatter}
\newpage

\section{Introduction}

The need for this work arose in the context of HERMES-TP/SP \footnote{High Energy Rapid Modular Ensemble of Satellites - Technological/Scientific Pathfinder} mission concept during 2017. HERMES aims to localize and study  bright high-energy transients---such as Gamma-Ray Bursts---by hosting miniaturized X/soft-$\gamma$ detectors on board nano-satellites in low-Earth orbit (LEO). 
HERMES detectors are designed around the so-called “siswich” concept \cite{Campana:2017jls} in which silicon detectors play the double role of sensor for the scintillation light emitted by suitable scintillator crystals and of 
independent detector for low energy X-rays.
The scintillator selected for use on HERMES units is GAGG:Ce (Gd$_3$Al$_2$Ga$_3$O$_{12}$:Ce, Cerium-doped Gadolinium Aluminium Gallium Garnet). It is a promising 
scintillation crystal displaying a wide array of appealing features for space applications: very high light-yield, fast-decay times, very low intrinsic background and mechanical robustness.\\
However, GAGG:Ce is still a recently developed scintillator \cite{KAMADA201288} and, as a consequence, literature is lacking on points crucial to its applicability in space. For example, GAGG:Ce is characterized by unusually intense and long-lasting afterglow emission \cite{LUCCHINI2016176}, a slow phosphorescence component in scintillation light. 
Afterglow emission is a source of background noise and is induced by the exposure of GAGG:Ce crystals to electromagnetic and particle radiation. 
Hence, in space applications, an effective degradation of the detector energy resolution should be expected as a result of the phosphorescence induced by the interaction of the energetic particles in the near-Earth radiation environment with the GAGG:Ce scintillators; the extent and dynamics of such phenomena depending on both the host spacecraft orbit and the crystal intrinsic properties. Besides degradation of the energy resolution, the current induced in the SDD sensors by the background light due to the afterglow emission may become too large, impairing the functionality of the Front-End Electronics (FEE).
To tackle this last concern we conducted an irradiation campaign at Trento Proton Therapy Center (TPTC) in which a GAGG:Ce sample was irradiated with 70 MeV protons. The choice of particle specie, energy and fluences was driven by the need to simulate the nature of the radiation environment of near-equatorial LEO orbits and the constraints of the cyclotron particle accelerator available at TPTC and of our equipment.\\ 
The corpus of this paper is arranged in three parts. In the first part we describe the experiment set-up and timeline. In the second part we discuss GAGG:Ce phosphorescence, introduce our model of the afterglow emission, discuss the impact of activation on our observations and outline the fit procedure and results. The afterglow model development is discussed in detail in \ref{sec:modelderivation}. Finally we make use of the afterglow model, supported by the AE9/AP9 trapped radiation belt models \cite{irenemodel}, to estimate the impact of the afterglow emission resulting from the orbital radiation environment on the performance of the instrument.

\section{Experiment Outline}
\label{sec:expout}

A GAGG:Ce scintillation crystal---dimensions $3 \times 1 \times 1$ cm$^3$---was housed in a lightproof metal case hosting two PMT detectors. 
The large faces of the crystal were wrapped in thin white teflon to minimize scintillation light dispersion, while both the unobstructed small faces were coupled to $5$ cm long quartz light guides by means of optical grade silicone grease. The same coupling technique was used to interface the light guides to the photocatodes of the Hamamatsu R4125 PMTs.
The anode signal of one of the two PMTs, labelled PMT1, was measured by a Keithley 6487 picoammeter. The signals from the last dynode of both PMTs were amplified by a factor of $20$, brought to a discriminator with thresholds set to $-50$ mV and redirected to a counter and a multi-channel digitizer for waveform acquisition, along with the anode signal of PMT2. Waveform acquisition could be triggered by either a single PMT or both, the latter through a programmable logic unit.
Irradiation took place at Trento Proton Therapy Centre (TPTC), Trento, Italy. At this facility a cyclotron (IBA, \textit{Proteus 235}) serves two medical treatment rooms and an experimental area, accelerating protons to a kinetic energy selectable in the range $70$--$228$ MeV. The proton beam has a Gaussian profile with $\sigma = 6.9$ mm at  $70$ MeV energy \cite{TPTC}. Two operation regimes are available: the so-called `dark' mode with very low beam intensity ($< 10$ protons s$^{-1}$), and the normal mode with high beam intensity. In normal mode, the extraction current is adjustable in the range $1$--$320$ nA and is modulated by a 50\% duty-cycle square wave with a period of 100 ms. We used both modes for different purposes. \\

\definecolor{color0}{HTML}{000000} 
\definecolor{color1}{HTML}{5c4ccf} 
\definecolor{color2}{HTML}{a3d4ff} 
\definecolor{color3}{HTML}{14b385} 
\definecolor{color4}{HTML}{a56bbf} 
\definecolor{color5}{HTML}{fa9507} 
\definecolor{color6}{HTML}{e84031}

\begin{figure}[htp]
\centering
\includegraphics[width=0.8\textwidth]{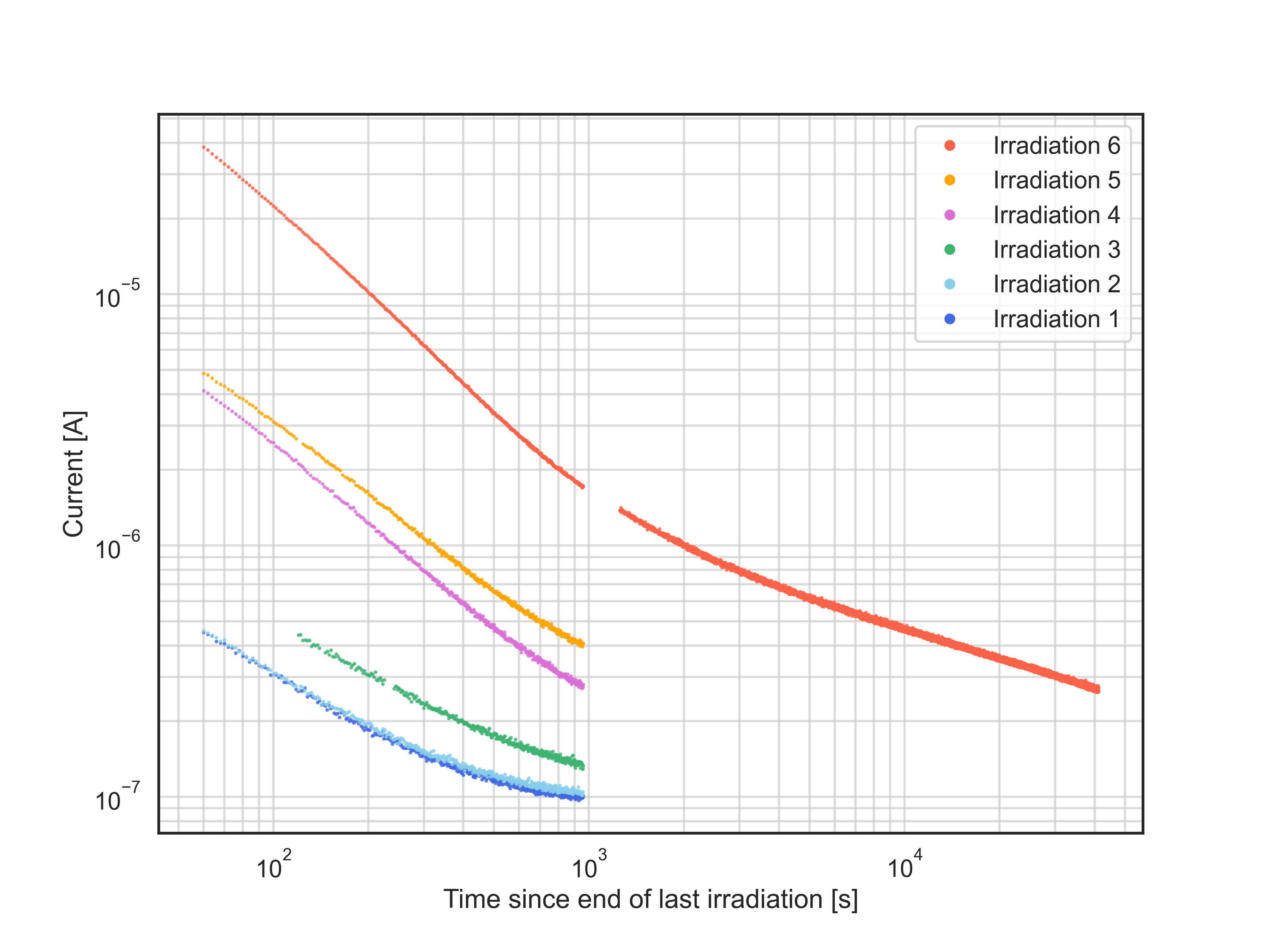}
\caption{PMT1 anode current induced by the afterglow resulting from irradiation with $70$ MeV protons of a GAGG:Ce sample, versus time elapsed from the irradiation end. Data from 30th January acquisitions. Color coding as in  Tab. \ref{tab:FULLdoses}.
The first $60$ s of data in the third measurement (green) have been excluded due to an improper initialization of the PMTs power supply.}
\label{fig:allrun}
\end{figure}

\begin{table}[h!]
\centering
 \begin{tabular}{| c c c c c c |} 
 \hline
 { } & {Ext. Current [nA]} & { \ Irr. Duration [s]} & { \ 
Intercepted protons [p]} & { \ Dose [rad]} & { \ E.O.A.$^{[1]}$} \\ [0.5ex] 
 \hline
 Irradiation 1
 \textcolor{color1}{$\bullet$} & 1 &  $90$ &  $(1.19 \pm 0.23)\times10^{8}$ & 6 & $1$y EQ\\
 Irradiation 2
 \textcolor{color2}{$\bullet$} & 1 &  $90$ & $(1.20 \pm 0.23)\times10^{8}$ & 6 & $2$y EQ\\
 Irradiation 3
 \textcolor{color3}{$\bullet$} & 1 &  $270$ & $(3.56 \pm 0.69)\times10^{8}$ & 19 & $5$y EQ\\
 Irradiation 4
 \textcolor{color4}{$\bullet$} & 10 &  $100$ & $(1.37 \pm 0.26)\times10^{9}$ & 73 & $10$y EQ\\
 Irradiation 5
 \textcolor{color5}{$\bullet$} & 10 &  $144$ & $(1.95 \pm 0.37)\times10^{9}$ & 104 & $2$y SSO\\
 Irradiation 6
 \textcolor{color6}{$\bullet$} & 100 &  $115$ & $(1.52 \pm 0.29)\times10^{10}$ & 807 & $10$y SSO\\[1ex] 
\hline

\end{tabular}

\caption{
Detailed table of the irradiation runs. Each row represents an irradiation step. Runs are identified by current log start time and, J2 irradiation steps are color coded as in the article body. The number of protons intercepted by the target and the total absorbed dose were estimated from a GEANT4 simulation of the experimental setup.
[1]: Equivalent Orbital Age. The reported values represent the number of consecutive years of orbits needed to the tested sample to achieve the same proton dose accumulated over different steps of the irradiation campaign. Computed according to AP8MIN models of the proton radiation environment (kinetic energies $> 0.1$ MeV) of an orbit with altitude $550$ km and inclination $10$ (EQ) or $98$ (SSO) degrees.
}

\label{tab:FULLdoses}

\end{table}
For our tests we selected a proton energy value of $70$ MeV. This choice was driven by the following reasons. First of all, the need of a pure proton beam with an energy representative of the trapped proton spectrum characteristic of low-Earth, nearly equatorial orbit. Most of the trapped protons in the regions of the inner Van Allen radiation belt near the equator have energies which spans tenths to hundreds MeV (see Fig. \ref{fig:intflux_cl90}). Moreover, $70$ MeV is the smallest energy attainable without further degradation at TPTC. Finally, simulations showed that $70$ MeV protons would most often release their whole kinetic energy within the GAGG:Ce crystal bulk.\\
Regarding the irradiation duration and the cyclotron extraction current selected for each irradiation step, hence the dose to be irradiated, we chose parameters such that the radiation dose accumulated by the GAGG:Ce target would match the end-of-life levels expected from different orbital operation scenarios. We remark on the $1$ nA beam minimum extraction current resulting in irradiation flux condition exceeding the levels expected in orbit by at least two order of magnitudes (compare Tab. \ref{tab:FULLdoses}, Fig. \ref{fig:intfluxmap} and Fig. \ref{fig:intflux_cl90}). Working with fluxes outside the expected conditions, possible saturation effects in afterglow emission may occur. Such phenomenon would result in an underestimate of the phosphorescence to be expected in orbit, where particle radiation fluxes are generally lower. This eventuality is addressed, at least in an approximate way, by our model of the crystal luminescence.\\
For each irradiation step the schedule of operations involved:
\begin{enumerate}
\item In dark mode: collection of $\sim 5000$ PMT2 anode current waveforms, triggered by the coincidence of the signals from both PMTs operated at $1100$ and $1350$ V (proton events).
\item In normal mode: irradiation of the crystal with the required beam intensity and exposure duration. During this step the PMTs were turned off.
\item Beam off: $60$ seconds after the end of each irradiation, measured by a chronometer, the anode current of PMT1, now operated at $1500$ V, was sampled each second for about $15$ minutes ($800$ seconds), see Fig. \ref{fig:allrun}. At the same time, count data were acquired both for individual PMTs last-dynode signal and their coincidence.
\end{enumerate}
Proton waveforms were acquired after the last irradiation step. Afterwards, monitoring of anode current was started once more and continued for $\sim 11$ hours during the night.
The scintillator temperature was monitored by means of a thermocouple and ranged between $21 \pm 0.5 \, ^\circ$C.
Specifications of each irradiation step are reported in detail in Tab. \ref{tab:FULLdoses}. 
The number of intercepted beam protons was estimated through GEANT4 simulation of a GAGG:Ce crystal irradiation experiment. 
In this simulation, a $70$ MeV proton beam with a 2D Gaussian profile of width $\sigma = 6.9$ mm  was modelled, according to the beam characteristics at the isocenter point \cite{TPTC}.  The crystal was placed at the beam center and possible positioning errors were taken into account. The total number of simulated protons was selected according to the flux measured by the beam monitor thus the computed intercepted fluence, total energy deposit and resulting dose.\\

\section{Afterglow emission models}
\label{sec:models}

\begin{figure}[htp]
\centering
\includegraphics[width=0.7\textwidth]{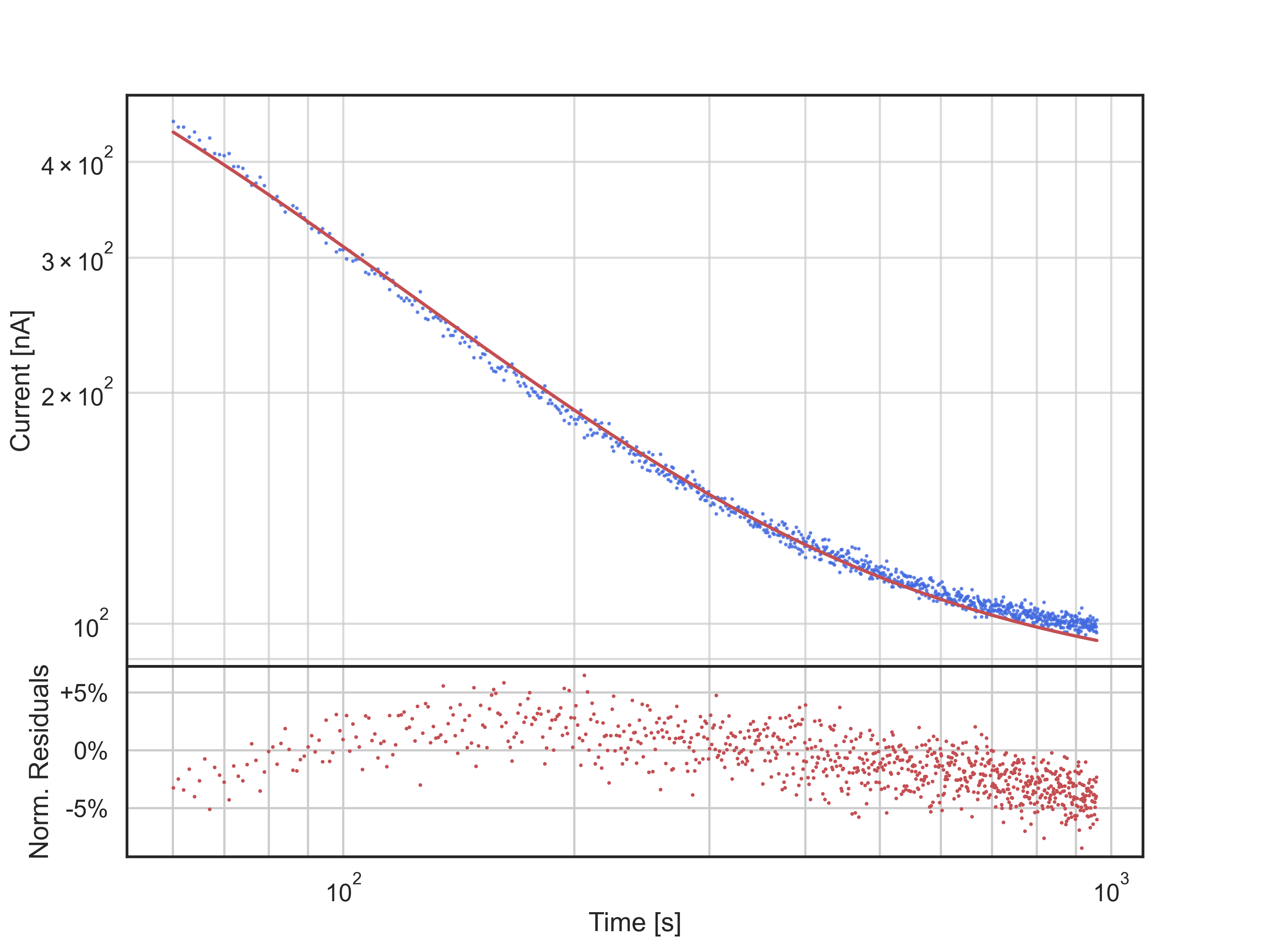}
\caption{Fit to inverse power-law plus a background constant of the PMT anode current from Irradiation 1, cf. Tab. \ref{tab:FULLdoses}. Best fit power-law index is $1.54$. Time-dependent patterns in residuals are apparent.}
\label{fig:powlawresidual}
\end{figure}

Long-lived afterglow emission in scintillators is attributed to the existence of intrinsic or impurity defects within the crystal lattice. Some of the charge carriers (electrons or holes) liberated by the ionizing radiation can be trapped at defect sites into metastable states. At later times, charge carriers  escape these sites by different processes (e.g. to the conduction band by thermal energy absorption \cite{knoll} or to nearby recombination centers by direct or thermally assisted tunneling \cite{2006:huntley}). Ultimately all of the charge carriers recombine, mainly through radiative paths in good scintillators, giving rise to luminescence.  Different scintillating materials display different afterglow characteristics. Although mitigation techniques (e.g. Mg-codoping) proved successful \cite{LUCCHINI2016176}, GAGG:Ce is known for its intense afterglow emission which may last up to several days \cite{Yoneyama_2018}.\\
In the existing literature GAGG:Ce afterglow emission has been reported decaying as a power-law of time \cite{LUCCHINI2016176}\cite{Yoneyama_2018}.
Although a power law can adequately describe the decaying emission within short times after the end of an exposure, we found such decay profile unable to fit the whole duration of our measurements. Deviations from power law behaviour are evident already by inspection of Fig. \ref{fig:allrun}. In Fig. \ref{fig:powlawresidual} we report on the power-law fit of one measure in our dataset.\\

In this work we seek a different approach to model the proton induced afterglow emission which allow us to accurately fit the data and to make predictions on the amount of afterglow to be expected in space applications.  
At the core of this representation lies the assumption that afterglow arises from the delayed recombination of charge carriers kept in metastable states by charge ‘traps’, which can be classified in species characterized by a mean \textit{capture rate} and a mean \textit{lifetime}. 
We expect that at the smallest time scales (seconds) accessible through our data the characteristic times of de-excitation of the metastable states do not form a continuum, so we assume there exist a discrete set of trap species. The dimension of this set is not known beforehand and has to be determined by data analysis.
At any given time the number of occupied traps can be determined balancing the rate at which new traps get occupied during irradiation versus the rate at which charges free themselves from traps to recombine. This mechanism is  described through a system of differential equations with form: 
\begin{linenomath}
\begin{equation}
 \frac{d}{dt}N_i(t) = n_i\,\phi - \frac{N_i(t)}{\tau_i} 
     \label{eq:occup_traps_time}
\end{equation}
\end{linenomath}
with $n_i$ and $\tau_i$ being the characteristic capture rate and mean lifetime of an $i$-labelled trap species. These equations can be analytically solved for arbitrary irradiation flux profiles $\phi$, hence for the irradiation timeline of our campaign.\\
This simple model is able to describe observations following the first four, low-dose, irradiation steps but fails when applied to the whole dataset. In order to support the observation which followed the fifth and sixth irradiation steps we modified the model to support linear variations in the traps capture rates.\\
A mathematical derivation of the model is given in \ref{sec:modelderivation}.\\

\section{Crystal activation}
\label{sec:activation}

The energies of the protons, both in the near-Earth radiation environment and in the irradiation campaign, are high enough to induce a level of activation in GAGG:Ce crystals.
In these settings, activation poses different challenges. In orbit, crystal activation is expected to result in a decay spectrum from the unstable nuclides that will interfere with the observation of astronomical gamma-ray sources. From the point of view of the irradiation tests, the scintillation accompanying the decays is superposed to the afterglow emission and the two contributions can not be disentangled from our observations.\\
The most direct (yet limited) information about the activation occurred during our irradiation campaign come from coincidence count data. A coincidence event was registered whenever current from both PMTs exceeded a pre-set signal threshold. Such events are more likely to result from the sudden, bright scintillation light  produced in nuclear decays than they are from the incoherent yet persistent afterglow emission.\\
The activation effects of proton irradiation on GAGG:Ce has been investigated  already for energy of $70$ MeV \cite{Yoneyama_2018} or higher \cite{sakano2014}. The work for Yoneyama et al., in particular, identified a number of lines in the GAGG:Ce activation spectrum with energies up to $1038$ keV resulting from a $10$ krad, $70$ MeV proton irradiation.\\
Using this information we estimated the contribution of scintillation to the measured currents at the beginning of each measurement (60 after the end of the corresponding irradiation step) assuming that on average each detected decay deposited 1 MeV of ionization energy in the crystal.
The decay rates needed to calculate the scintillation component were estimated from the coincidence count curves by fitting them to a simplified model that considers one of the decay modes to be dominant.
Taking into account our estimate of light transport efficiency, PMT gain and the picoammetter integration time, analysis of count data from our campaign lead to an estimated contribution to the PMT anode current from activation at the beginning of the measurements following the first irradiation equal to $25$ nA, or $6.7\%$ of the current value observed at that time. The same reasoning, applied to the data observed after the last, high-flux step of irradiation, lead to a slightly larger ratio estimate of $8.2\%$.
Far from being conclusive, this analysis shows that the contribution from activation to the measured PMT anode current was not negligible.\\

\section{Fitting data to the afterglow emission model}
\label{sec:fit}

Due to a significant presence of scintillation light in the measurements the application of the afterglow models we developed is not justified because it entails precise meaning to the fit parameters.
Nonetheless, being interested in a conservative estimation of the afterglow emission during the mission, the additional component due to crystal activation can be considered as yet another source of overestimation and we can still apply our modelling in a purely empirical way to reproduce the light curves, provided we do not attach any precise meaning to the model parameters.
For this reason we choose to assign new names to the parameters, according to the following scheme: instead of trap species we talk about emission modes, the capture rates become emission amplitudes, and the variation of the capture rates turn into the variation of the emission amplitudes.\\

The PMT used in our measurements intercepted only a fraction of the photon flux emitted by the crystal.
The anode current $I(t)$, as measured by the picoammeter, can be expressed as:
\begin{linenomath}
\begin{equation}
    I(t) = e G_V f_q \phi_q(t)
    \label{eq:current}
\end{equation}
\end{linenomath}
where $e$ is the elementary charge, $G_V$ is the gain of the PMT at operational voltage $V$, $f_q$ is the fraction of photons that extract an electron from the PMT photocathode, and $\phi_q(t)$ is the photon flux from Eq. \ref{eq:photon_flux}.\\
Through PMT calibration
we found a gain value $G_{1500} = (1.04 \pm 0.04)\cdot 10^6$. The quantity $f_q$ depends on both the wavelength dependence of the light transport efficiency from the crystal to the PMT and the PMT spectral response \cite{hamamatsu_tubes}. It can be expressed as the average of the product of these efficiencies, weighted by the GAGG:Ce emission spectrum \cite{Yoneyama_2018}:
\begin{linenomath}
\begin{equation*}
f_q = \langle \epsilon_{LG}(\lambda) \epsilon_{PMT}(\lambda) \rangle_{GAGG:Ce}
\end{equation*}
\end{linenomath}
In our analysis we assumed that $\epsilon_{LG}$ had a negligible variation over the relevant wavelength interval. Therefore it is possible to decouple  contributions from the transport efficiency and the PMT response:
\begin{linenomath}
\begin{equation*}
 f_q = \epsilon_{LG} \epsilon_{PMT} = \epsilon_{LG} \langle \epsilon_{PMT}(\lambda) \rangle_{GAGG:Ce} \;\text{.}
\end{equation*}
\end{linenomath}
The average PMT quantum efficiency was calculated using data in the literature to be $\epsilon_{PMT} = 0.06 \pm 0.01$ \cite{hamamatsu_tubes}.\\
The fraction of light emitted by the crystal reaching the PMT photocathode was obtained comparing the photons collected by the PMT with the scintillation light emission expected from irradiation of GAGG:Ce with Cs-137: 
\begin{linenomath}
\begin{equation*}
\epsilon_{LG} = \frac{Q_\gamma}{e \epsilon_{PMT} G_V E_\gamma}\bigg/LY
\end{equation*}
\end{linenomath}
where $Q_\gamma$ is the integral of the current pulses corresponding to the gamma ray photons of energy $E_\gamma$, and $LY = 53 \pm 3$ photons/keV is the light-yield of the crystal measured by its manufacturer \cite{gaggJ2}. For the Cs-137 $662$ keV line we found that PMT1 collected $15 \pm 4$ photons/keV, leading to an estimate of the light transport efficiency $\epsilon_{LG} = 0.29 \pm 0.09$.\\

\begin{figure}[htp]
\centering
\includegraphics[width=0.7\textwidth]{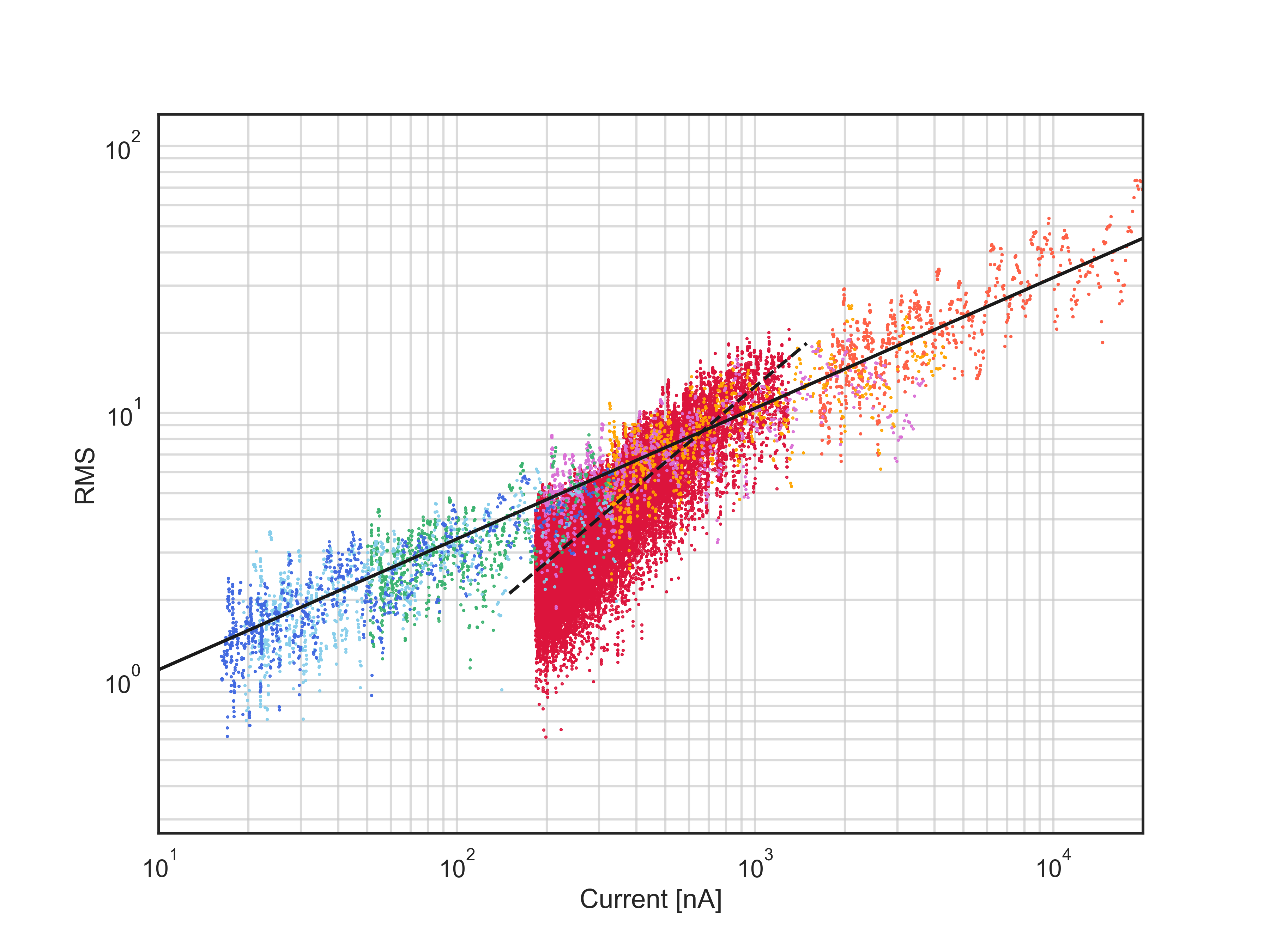}
\caption{RMS values computed from the residuals of cubic spline fits of the light curves. The set in red behaving differently corresponds to the night measurement of the afterglow following the last irradiation. The RMS values were modelled as $RMS = A+B\cdot I^C$. Best fit to this law is shown in black solid line for the measurements immediately following the irradiation steps, and refers to the values: $A = 0$, $B = 0.35$, and $C = 0.49$. The black dashed line represents the best fit of the night RMS dataset with parameters $A = 0$, $B = 0.019$, and $C = 0.94$. See Tab. \ref{tab:FULLdoses} for color legend.}
\label{fig:rms}
\end{figure}

To fit the model to the observations we first estimated the uncertainties associated to each measurement, which we assumed to be well represented by the fluctuations in the data. This  procedure has been accomplished in two steps. In the first we smoothed the light curves by fitting them with cubic splines and then recovered the fluctuation by subtracting the smoothed curves from the observations. In the second step we determined how the fluctuations changed as a function of the intensity of the light reaching the PMT. We performed a calculation of the moving RMS of the fluctuations, with different window sizes to determine its robustness. The RMS values with a window size of $8$ samples are plotted in Fig. \ref{fig:rms} as a function of the smoothed PMT anode current, to which we subtracted our best estimate of the dark current ($83$ nA). We fitted the RMS values to a power law and a residual constant term:
\begin{linenomath}
\begin{equation*}
RMS = A+B\cdot I^C
\end{equation*}
\end{linenomath}
For the measurements immediately following each irradiation step we found that the RMS changed in a way approximately proportional to the square root of the light flux, while in the night measurement the decaying emission was followed linearly.\\

The fit of the observation dataset to the afterglow models was performed in an iterative way, progressively adding the data of the irradiation steps to see which changes in the model were needed to adequately describe the observations. 
We found that a model with constant emission amplitudes is able to reproduce the dataset only up to the fouth irradiation step.
We looked for evidences of radiation damage analyzing the scintillators light-output before and after the irradiations. 
In fact, radiation damage is expected to affect light-output in scintillators by reducing trasmittance through creation of color centers absorbing scintillation light \cite{knoll}. For GAGG:Ce the color centres absorption band appears to be in the UV range \cite{alenkov2019}. Given the small overlap between the absorption spectrum and the afterglow emission spectrum, it is difficult however to appreciate evidences for this phenomenon from our observations, hence to unambiguously attribute a variation in the emission amplitudes to radiation damage. From the analysis of the proton event waveforms we were unable to infer any significant ($> 5\%$) reduction in light-output.
Before addressing the complete observation set we performed a fit of the measurements taken following the last irradiation, both immediately after and during the night, to the model with constant emission amplitudes, taking into account all the irradiation steps. 
We repeated the fit several times to determine the number of emission modes, and we found an optimal value of seven: with fewer modes evident patterns remained in the fit residuals, while a larger number of emission modes resulted in worse estimation of the best fit parameters without significant improvements in the residuals.\\ 
\begin{figure}[htp]
\centering
\includegraphics[width=0.8\textwidth]{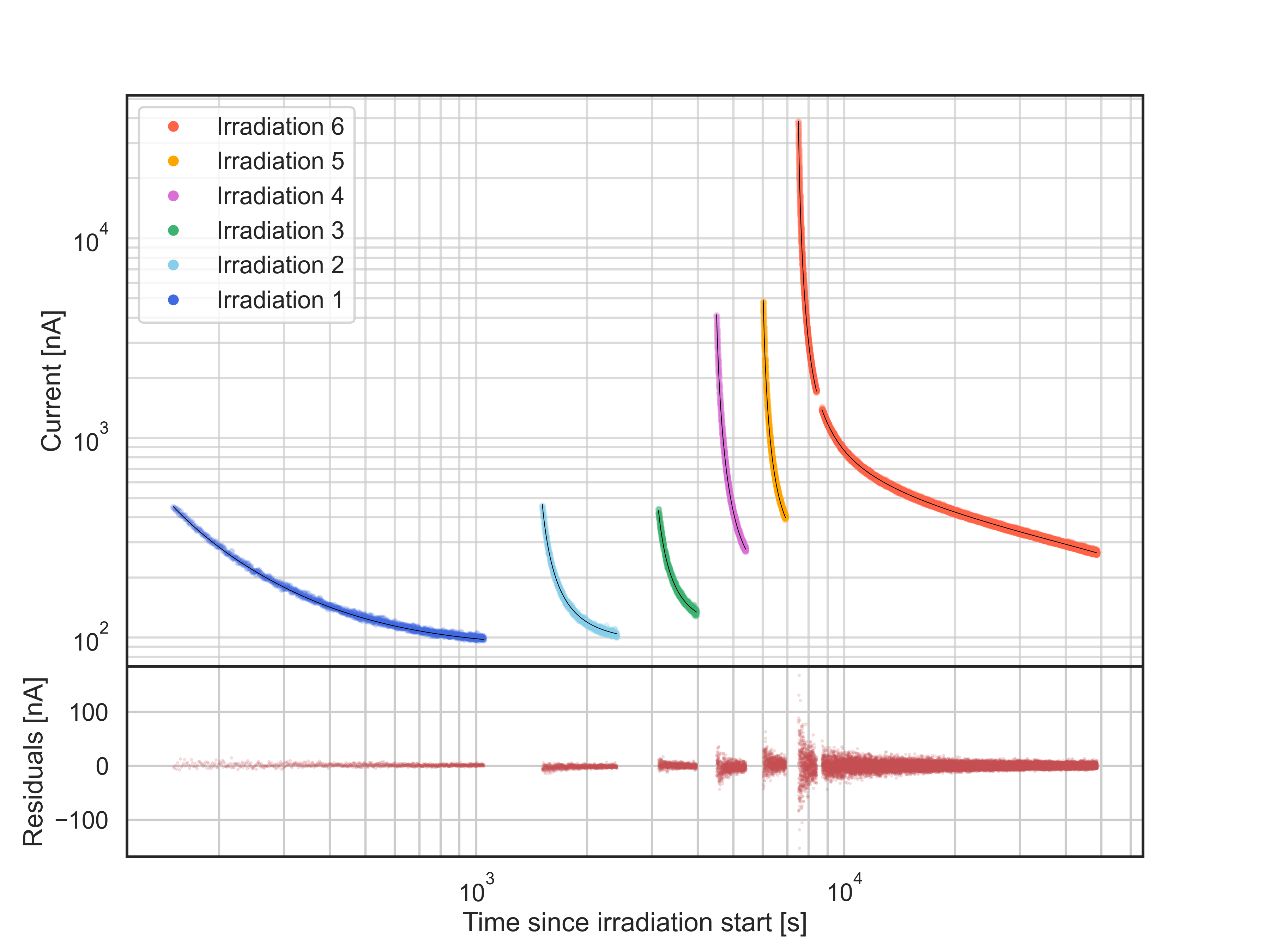}
\caption{Best fit of the January 30th dataset to the model of Sec. \ref{sec:models}. Parameters estimates are reported in Tab. \ref{tab:finfit_nt}. Temperatures in range $21 \pm 0.5 \, ^\circ$C} 
\label{fig:finfit}
\end{figure}
Fit results are graphically reported in Fig. \ref{fig:finfit}, while the best fit parameters values are reported in Tab. \ref{tab:finfit_nt}.\\
\begin{table}[htp]
\centering
    \begin{minipage}{.6\textwidth}
      \centering
		\begin{tabular}{cccccc}
		\toprule
		$\tau_i$ [$s$] & $\sigma_\tau$ [$s$] & $n_i$& $\sigma_n$ & $\Delta n$ & $\sigma_{\Delta n}$\\
		\midrule

		$24.9$  & $0.8$ & $197.1$   & $80.9$   & $72.0$    & $31.5$\\
		$69.1$  & $2.8$ & $121.1$ & $49.4$ & $-27.6$ & $12.5$\\
		$194.1$ & $4.3$ & $93.4$  & $38.2$ & $-30.6$ & $13.4$\\
		$697.1$   & $29.7$  & $48.7$  & $20.2$ & $-27.3$ & $12.9$\\
		$2239.5$  & $106.2$  & $35.0$  & $14.8$ & $-18.7$ & $10.9$\\
		$8315.7$  & $242.8$ & $44.8$  & $18.2$ & N/A     & N/A\\
		$70138.5$ & $937.5$ & $375.5$ & $152.8$ & N/A     & N/A\\

        \midrule
		\end{tabular}
		\caption{Emission mode time constants ($\tau$), amplitudes ($n$) and their relative variations ($\Delta n$) as estimated from fit to the models of Sec. \ref{sec:models}. The amplitude value is fixed for the last two emission modes, while it changes proportionally to the fluence for the other modes. The uncertainties include all the known sources of error, see the text. Afterglow data from  GAGG:Ce sample at temperature $21 \pm 0.5 \, ^\circ$C.}
		\label{tab:finfit_nt}
    \end{minipage}
\end{table}

The information about emission modes with time constants larger than about a hour comes almost exclusively from the night measurement. For this reason we assigned constant amplitudes to these emission modes.
The uncertainties in the the model parameters obtained through the fit procedure are only due to the fluctuations in the data, since we replaced all the other quantities in Eq. \ref{eq:current} and the irradiation fluxes with their most probable values. Both the fraction of emitted photons that extract an electron at PMT1 cathode and the fraction of the proton beam intercepted by the crystal are constant multiplicative quantities for which the error propagation is straightforward. 
However, the model is non-linear in the fluences and gain parameters. Hence a different approach was needed to estimate the contributions to uncertainties from these quantities. We used Monte Carlo techniques, starting from the estimated uncertainties, to evaluate the variation of the model parameters subject to compatibility with the measured data. Results are summarized in Tab. \ref{tab:finfit_nt}.\\

\section{In-orbit impact of GAGG:Ce afterglow on silicon drift detectors}
\label{sec:leakage}

\begin{figure}[htp]
\centering
\begin{subfigure}{.45\textwidth}
\includegraphics[width=1\textwidth]{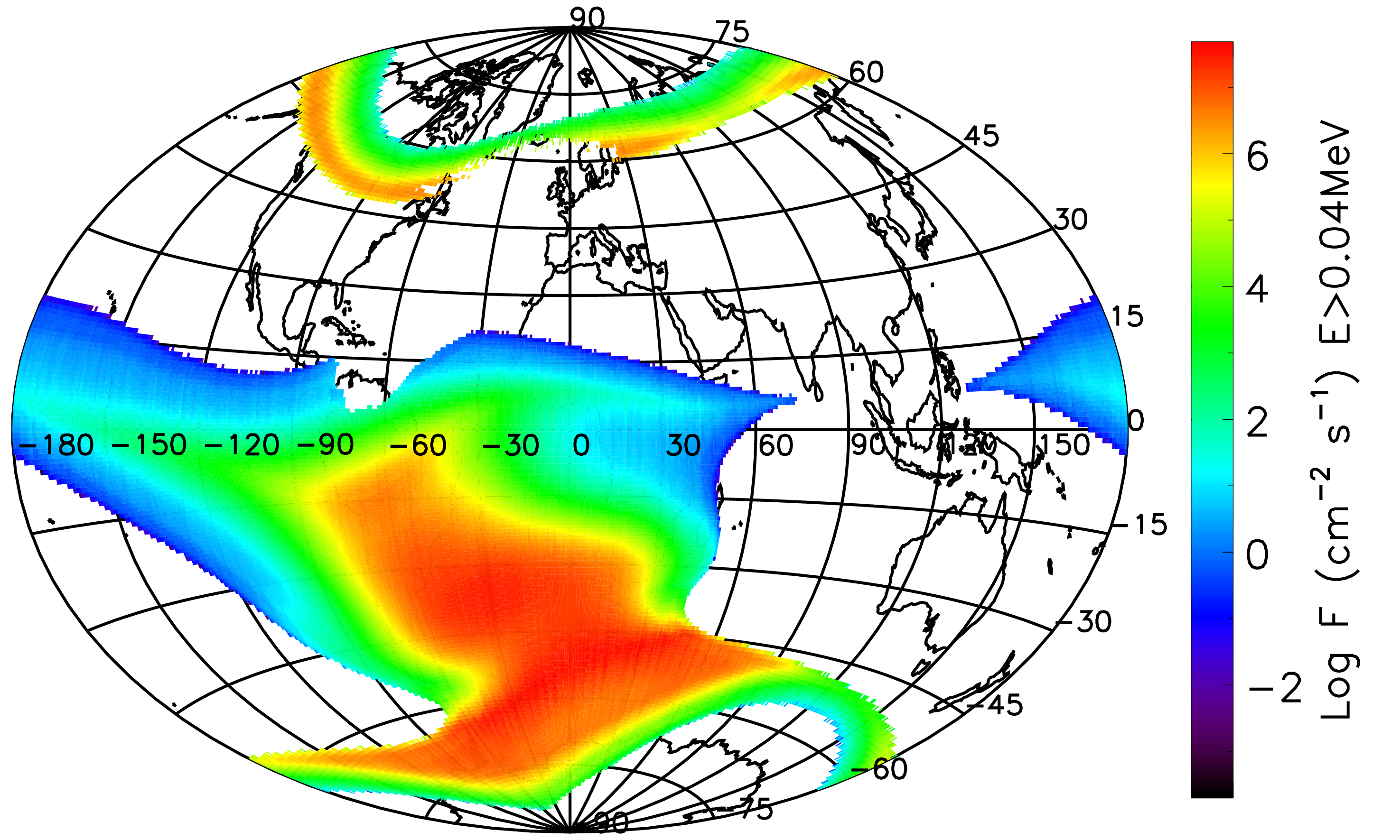}
\caption{Trapped electrons with kinetic energy exceeding $0.04$ MeV.}
\end{subfigure}
\begin{subfigure}{.45\textwidth}
\includegraphics[width=1\textwidth]{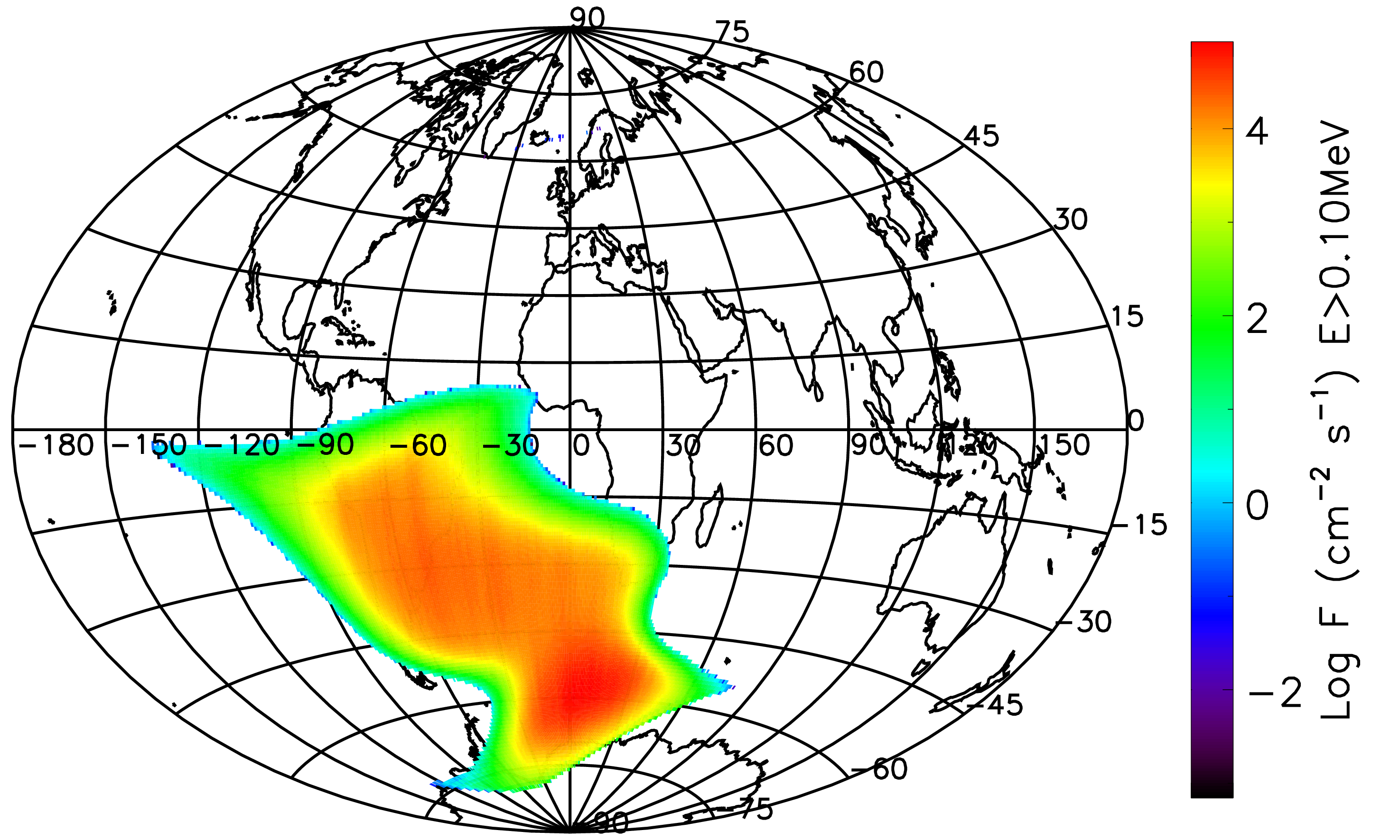}
\caption{Trapped protons with kinetic energy exceeding $0.1$ MeV.} 
\end{subfigure}
\caption{Integral flux maps. Results obtained from IRENE AE9/AP9 model in pertubed Monte Carlo mode, at $90\%$ confidence level for $100$ different iterations.}
\label{fig:intfluxmap}
\end{figure}

\begin{figure}[htp]
\centering
\includegraphics[width=0.7\textwidth]{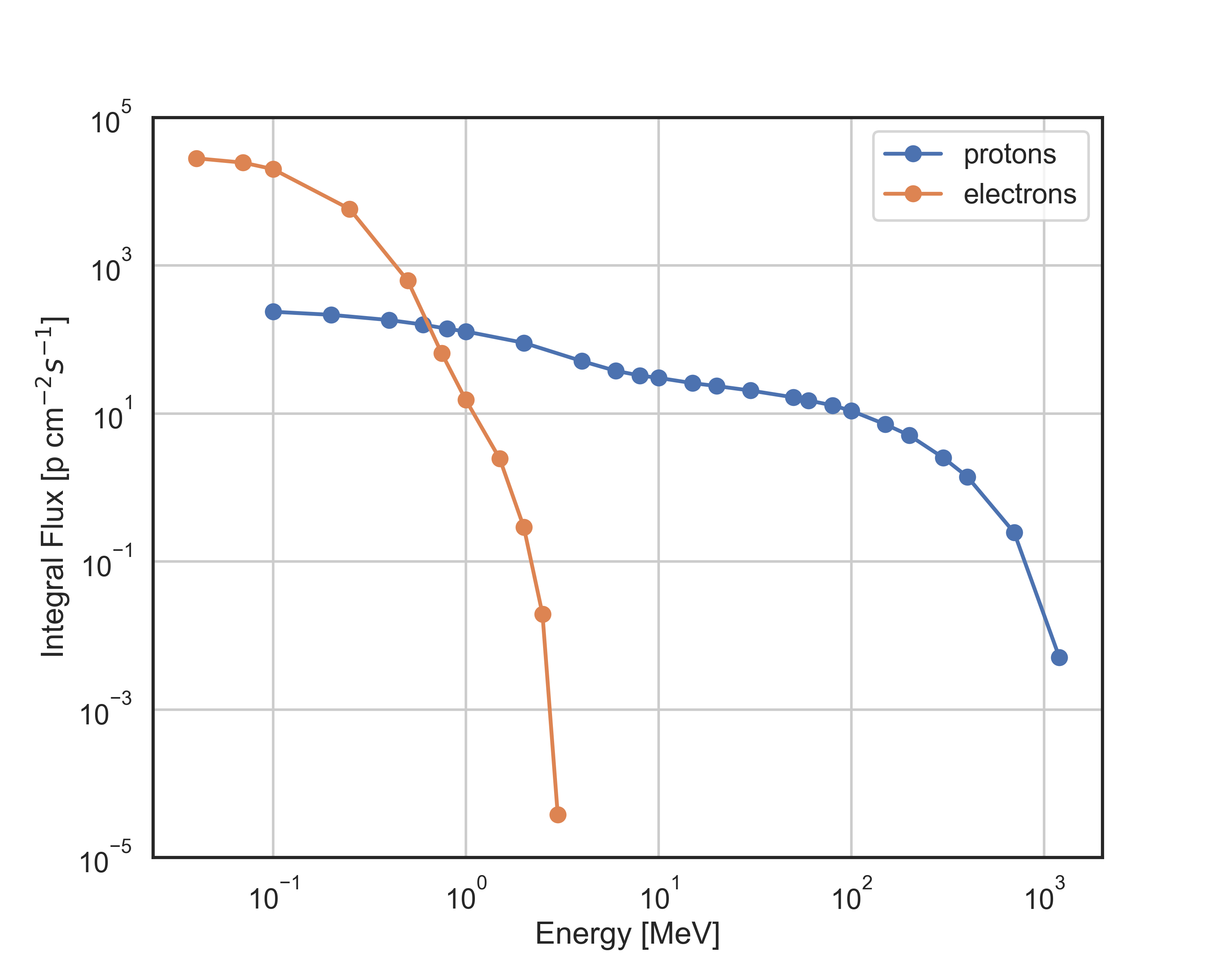}
\caption{Proton and electron average integral flux for an orbit with altitude $550$ km and inclination $10$ degrees. Results obtained from IRENE AE9/AP9 model in pertubed Monte Carlo mode, at $90\%$ confidence level for $100$ different iterations.} 
\label{fig:intflux_cl90}
\end{figure}

On average the GAGG:Ce afterglow emission manifests itself as a continuous stream of optical photons with monotonically decreasing flux after stimulation. 
The randomly arriving photons are able to induce detectable anode current pulses on PMTs but not on the highly-efficient yet unamplifying SDDs. Hence scintillator afterglow will not result in triggering HERMES SDD front-end electronics. Instead it will behave as an equivalent leakage current component adding to the true device current.\\
In space most of the afterglow emission will be induced by the the interactions of the particles trapped in the Van Allen radiation belts and the scintillator material. At LEO altitudes most of the trapped particles are concentrated in the South Atlantic Anomaly (SAA) and the polar regions. The HERMES foreseen orbit is at low inclination, thus spanning a restricted range of geomagnetic latitudes and grazing the South Atlantic Anomaly in its outermost regions \cite{campana2020}.\\
Afterglow emission generated during SAA fly-overs is expected to lead to a periodic modulation in device current, thus a periodic degradation of the HERMES detector energy resolution. Most importantly the total device current may exceed the maximum value the SDDs FEE is capable to cope with. 
The latter contingency is investigated in the remainder of this paper through the application of the model described in Sec. \ref{sec:models} and
Sec. \ref{sec:fit} and the solution of  Eq. \ref{eq:occup_traps_time} for irradiation flux profiles representive of those to be expected during space operations.\\
In order to calculate the trapped particle fluxes expected along the orbit we used the IRENE (International Radiation Environment Near Earth) AE9/AP9 models \cite{irenemodel}. These are empirical models for computing proton and electron orbital fluxes. 
AE8/AP8 versions were developed by NASA and are regarded as the standard tool for radiation belt modeling \cite{misc:esa_models}. 
In the near future, the IRENE AE9/AP9 models---which are built upon much more recent trapped radiation observations---are expected to fully replace their predecessors. 
Both models have been shown to be in disagreement with recent trapped particle radiation observations in low-altitude, low-inclination orbits. In particular, in-situ measurements of trapped particles fluxes have been found to sit in between the predictions of the two models for these orbits, with AE8/AP8 underestimating and AE9/AP9 overestimating particles fluxes \cite{2014:campanasax} \cite{ripa2020}.\\ 
The AE9/AP9 models were run in Monte Carlo mode. In this mode the flux data contains all of the perturbed mean uncertainty plus an estimate of the variations due to space weather processes. In the Perturbed and Monte Carlo modes, the user sets a number of iterations and has the choice to compute the aggregated mean or percentile across the different runs at each time-step \cite{irenemodel}. The results of this work were obtained considering $100$ iterations at $90\%$ confidence level with time-step $10$~s and duration $30$~days for an orbit with altitude $550$ km and inclination $10$ degrees, coherently with the HERMES-TP/SP mission profile. The resulting average integral flux of trapped protons and electrons are represented in Fig. \ref{fig:intflux_cl90}. In Fig. \ref{fig:intfluxmap} the integral flux maps of trapped particles at altitude $550$ km are shown, as expected according to the IRENE AE9/AP9 models.\\

The HERMES detector will accomodate $60$ GAGG:Ce scintillators, each of dimensions $12.10 \times 6.94 \times 14.50$ mm$^3$. We considered a single crystal of the same dimensions. 
All the crystal surfaces were supposed to be reflective, except for one small face coupled to a pair of SDD cells covering the entire face of the crystal, each of effective area $6.94 \times 6.05$ mm$^2$. Considering the crystal completely shielded on the side opposite to the SDDs, the particle fluxes were integrated upon the remaining five faces of the scintillator.\\
Since the afterglow model was developed to explain data acquired for a GAGG:Ce sample at temperature $21 \pm 0.5 \, ^\circ$C, the results of this section refer to a crystal in the same temperature range. The in-orbit scintillators temperature is expected to be lower, ranging between $-20^\circ$ and $0^\circ$. We expect the intensity of the afterglow emission to decrease with the temperature as a consequence of the increased mobility of the charge carriers. \\
We used the model of Sec. \ref{sec:fit} to estimate the leakage current induced by afterglow emission on a single SDD. In particular, referring to Eq. \ref{eq:photon_flux} and \ref{eq:current}, the expected value of the leakage current is:
\begin{linenomath}
\begin{equation}
I_L(t) = e\, f\, \epsilon\, \phi_q(t)
\label{eq:leak}
\end{equation}
\end{linenomath}
Where $\epsilon$ indicates the quantum efficiency of the SDD, $f$ is the photon transport efficiency from crystal to SDD and $e$ is the elementary charge.
The SDD quantum efficiency $\epsilon$ was assumed to be constant at value $1$ over the afterglow emission spectrum. The photon transport efficiency from the crystal to an SDD cell was set to $0.5$ since on average the two cells should receive an equal number of photons. Considering the scintillator as continuosly irradiated at constant average rate for periods of duration equal to the orbital simulation time-step and taking into account the flux values calculated with AE9/AP9, we estimated $\phi_q(t)$ according to a worst-case parametrization of the afterglow emission in which the emission modes have constant amplitudes:
\begin{linenomath}
\begin{equation*}
n_{i}^{w.c.} = \max{(n_i, n_i + \Delta  n_i)}
\end{equation*}
\end{linenomath}
where $n_i$ and $\Delta n_i$ are the best fit values reported in Tab. \ref{tab:finfit_nt}.\\
Instead of using the ionization energy loss of the particles in the crystal, we considered their kinetic energy and normalized the differential fluxes of trapped particles to the kinetic energy of the protons in the irradiation campaign, i.e. we scaled the flux of particles at energy E (expressed in MeV) by 70/E \footnote{For example, a $70$ p cm$^{-2}$ s$^{-1}$ flux of $1$ MeV particles is converted to a $1$ p cm$^{-2}$ s$^{-1}$ flux of $70$ MeV protons}. 
Since more energetic protons are not stopped inside the crystal, scaling was performed only for kinetic energies lower than $70$~MeV.  Particles with energies over $70$~MeV were assigned an energy deposit equivalent to the one of a $70$~MeV proton. \\
\begin{figure}[htp]
\centering
\includegraphics[width=0.8\textwidth]{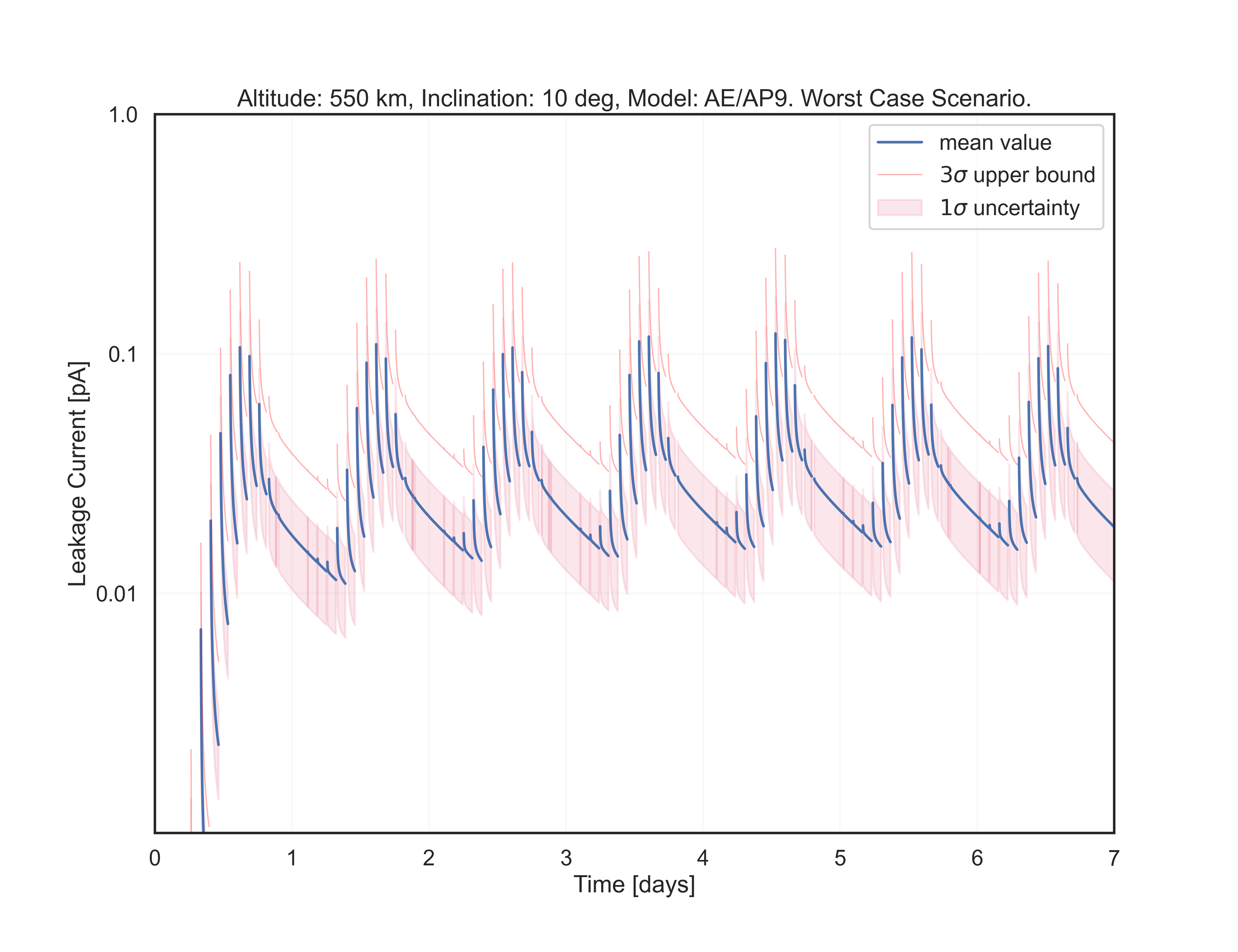}
\caption{Estimated worst-case leakage current of an SDD cell of dimension $6.94 \times 6.05$ mm$^2$ induced by GAGG:Ce afterglow emission, as expected from orbital irradiation of a $12.1 \times 6.94 \times 14.50$ mm$^3$ scintillator  at a temperature of $21 \pm 0.5 \, ^\circ$C in $550$ km, $10$ degrees orbit over $\sim 7$ days period ($100$ orbits). 
Values during transits over SAA were not computed.
The $1\sigma$ uncertainty region is represented as a shaded band. The $3\sigma$ upper-bound is reported as a red solid line. Orbital populations of protons and electrons were modelled through AE9/AP9 packages.} 
\label{fig:leakage_worstcase}
\end{figure}
In Fig. \ref{fig:leakage_worstcase} we report the resulting estimate of the equivalent leakage current induced by GAGG:Ce afterglow on the SDD, as expected over $100$ orbits at altitude $550$ km. 
We do not display values inside and up to one minutes after SAA transits. This choice is due to our model being not valid at times earlier than one minute after the end of an irradiation. Moreover HERMES instruments will be turned off during SAA fly-over.
The reported minimum, mean and maximum values of leakage current were computed starting after $24$ hours of orbital lifetime. 
Given the exponential nature of the afterglow model, these numbers are equally well representative when a larger number of orbits are taken into consideration. For instance, considering the full $30$ days simulation ($\sim 400$ orbits) we found an increase in the computed maximum leakage current of about $2\%$, which is comparable to the geographical fluctuations of trapped particle fluxes on the same interval.
In our calculations the increase in leakage current due to displacement damage in the SDDs was not taken into account.\\
The HERMES low-noise front-end electronics (FEE) is able to grant nominal performance up to $\sim 100$ pA of leakage current, a value well above the estimated maximum (by about two orders of magnitude). We conclude that GAGG:Ce afterglow should not endanger the well-functioning of the FEE.
However, when the HERMES fleet will be enlarged---eventually hosting spacecrafts in orbits at higher inclinations and delving in regions of much higher trapped particles concentration---the impact of afterglow on detector performance will demand further investigation.\\

\section*{Conclusions}
In this paper we discussed our investigation of GAGG:Ce for applications in the context of the HERMES-TP/SP nanosatellite mission. The goal was to determine whether the delayed luminescence caused by the interaction between the scintillators and the energetic particles of the near-Earth radiation environment could pose a threat to the well-functioning of the detector. For this reason, we conducted an irradiation campaign in which a GAGG:Ce sample was irradiated with $70$~MeV protons at dose levels representative of those expected from space operations. To translate our observations into predictions, we developed a model of GAGG:Ce afterglow and used the IRENE AE9/AP9 software package to describe the radiation environment in which the first HERMES spacecrafts will operate. We concluded that GAGG:Ce afterglow emission should not endanger the HERMES TP/SP operations.\\
Albeit being used in a pure empirical way, due to a significant presence of scintillation from crystal activation, the afterglow model we built has proven successful in describing our observations.  However uncertainties remains about dose-dependent variations in the afterglow decay and the emission behaviour at time scales larger than one hour. To further investigate these points, we are preparing a new, week long GAGG:Ce irradiation campaign.

\section*{Acknowledgments}
This study was funded by the European Union Horizon 2020 Research and Innovation Framework Programme under  the grant agreement HERMES-Scientific Pathfinder n. 821896 and from ASI-INAF Accordo Attuativo HERMES Technologic Pathfinder n. 2018-10-HH.0. The research leading to these results has also received funding from the European Union’s Horizon 2020 Programme under the AHEAD2020 project grant agreement n. 871158. We kindly thank the ReDSoX collaboration, the TIFPA staff and the anonymous reviewers for their precious contributions.

\clearpage
\appendix
\renewcommand{\thefigure}{A\arabic{figure}}
\setcounter{figure}{0}

\section{Detailed model derivation}
\label{sec:modelderivation}

To build a viable model of the radiation induced afterglow we made some simplifying assumptions motivated by the fact that our observations are largely unable to account for the details of the physical processes at play.
\begin{enumerate}
    \item Since we started observing the luminescence one minute after the end of each irradiation, and we sampled the PMT anode current once every second, our measurements are not sensitive to features in the signals that change at time scales shorter than few seconds. We expect that at the time scales accessible through our data the characteristic times of de-excitation of the metastable states do not form a continuum, so we assume there exist a discrete set of mean lifetimes of the occupied metastable states that we call ‘trap species’. The dimension of this set, $N_t$, is not known beforehand: it has to be determined by data analysis.
    \item The electrons in the ionization cloud produced by the interaction of a proton recombine or get trapped in a time span that is much shorter than the minimum time scale our experiment is sensitive to. For this reason we simplify the description of the capture process considering it to be instantaneous, and by assigning a given, yet unknown, discrete probability distribution for the electrons to be captured in the trap species. In principle this probability distribution could change for the ionization clouds of different protons because of statistical fluctuations in ionization, uneven trap distributions, and so on.
    \item We assume that the distributions of the trap species are uniform, or at least vary slowly, within the crystal volume.
    \item We assume that the densities of the trap species are much higher than the largest density of electrons in the conduction band, so that the capture probabilities have at most a very week dependence on the number of occupied traps, which we initially neglect.
    \item We ignore retrapping of the electrons emitted by the metastable state. Indeed, as will verified by the results of the afterglow data analysis, the probability for the electrons to be trapped in the relevant metastable states is so small that our measurements are not sensitive enough to be affected by this phenomenon.
    \item We ignore the temporal structure of the beam because it has a negligible impact: the correction factor on the emission intensity is less than $1$\% for mean lifetimes larger than $2.5$ s. 
\end{enumerate}
A further simplification stems from the large number of ionization electron-hole pairs liberated by a single proton (several millions), and from the large number of protons striking the crystal within the time resolution of our measurements: these large numbers allow us to mathematically threat the problem using continuous variables and differential equations, by employing averaged quantities, releasing ourselves from keeping track of the fluctuations involved in the real physical processes. This is also true when considering light excitation of the crystal.\\

In general, we can consider the capture process of an electron by a trap, or a recombination center, governed by an elementary probability $\pi$ which may differ according to the trap or recombination center species. Let us call $d$ their densities, and assign indices $r$ to recombination centers, $k$ to the trap species with mean lifetimes so short that they are not observable in the experiment, and $i$ to the trap species we are interested in. It can be shown that if the densities are constant in the volume of the crystal, besides edge effects, the capture probability of the $i$-labelled trap species can be written
\begin{equation}
    p_i = \frac{\pi_i d_i}{\sum_r\pi_r d_r+\sum_k\pi_k d_k+\sum_i\pi_i d_i}
    \label{eq:prob}
\end{equation}
In a good scintillator most of the electrons recombine radiatively, so $p_i \ll 1$. This is the reason why we neglect recapture of the charge carriers emitted by one of the traps into the traps we are interested to. In the case in which the densities of recombination centers and traps vary slowly within the volume of the crystal we replace Eq. \ref{eq:prob} with its average over the volume of the crystal affected by ionization.\\
As a way to further simplify the calculation, we relate our averaged quantities to a single incident particle since, from a practical point of view, they all produce the same average ionization. This choice allows for a simple rescaling of the results by the ratio of the average ionization energies when we apply our model to the prediction of the afterglow expected in the space radiation environment of the HERMES nano-satellites.\\
Since we initially consider the capture probabilities to be constant, we can treat the electron capture by the different trap species independently. Focusing our attention to the $i$-labelled species, then, the number $N_i$ of occupied traps varies during a given constant stimulation (either irradiation or illumination) according to the following differential equation:
$$ \frac{d}{dt}N_i(t) = n_i\,\phi - \frac{N_i(t)}{\tau_i} $$
The first term on the RHS accounts for the rate of electrons being trapped during irradiation, which is given by the product of the radiation flux integrated over the exposed surface of the crystal, $\phi$, with the average number of electrons trapped for particle of incident radiation, $n_i$. This last quantity is directly linked with measurement data, and it may be used to calculate approximate average capture probabilities of the trap species by dividing its value by the average number of ionization electron-hole pairs produced by an incident particle. For simplicity we will often refer to $n_i$ as the ‘capture rate’ of the traps.\\
The second term accounts for the electron emission from the traps, which happens with a rate that is proportional to the reciprocal of their mean lifetime in this particular metastable state.\\
The general solution to the equation above, referred to the $j$-th stimulation in a set, is
\begin{equation}
    N_i(t) = N_{i,j}^0 e^{-\frac{t}{\tau_i}} + n_i \phi_j \tau_i \left(1 - e^{-\frac{t}{\tau_i}}\right)
    \label{eq:model_res}
\end{equation}
where the time origin is at the beginning of the stimulation, and $N_{i,j}^0$ represents the number of traps already occupied at $t = 0$. Since the system is linear, we can treat separately the initial condition and the effects of all the $m$ stimulation steps, thus the change in occupied traps at the end of the $j$-th stimulation, after time $\Delta t_j$, due only to the excitation it provided, is
\begin{equation}
    \Delta N_{i,j} = n_i\tau_i\phi_j \left(1-e^{-\frac{\Delta t_j}{\tau_i}} \right)
    \label{eq:DN_fixed}
\end{equation}
By introducing $t_j$, the time at which the $j$-th stimulation starts, and $t_{>j} = t-t_j-\Delta t_j$, the time referred to the end of the $j$-th stimulation, the number of occupied states of the trap species we are considering evolves with time, valid outside the stimulation steps, in the following way:
\begin{equation}
    N_i(t) = N_i^0\;e^{-\frac{t}{\tau_i}}+\sum_{j=1}^m \theta(t_{>j})\; \Delta N_{i,j}\; e^{-\frac{t_{>j}}{\tau_i}}
    \label{eq:trapag}
\end{equation}
where $\theta(x)$ is the Heaviside step function.\\
The afterglow model developed so far can be used to describe the emission of the crystal after a low intensity stimulation if radiation damage can be neglected. Under intense excitation conditions, some of the trap species may fill up rapidly, so their capture probabilities will decrease, and other traps may become more efficient in trapping the electrons. Indeed, in Eq. \ref{eq:prob} the parameters $d_k$ and $d_i$ should refer to the densities of the unoccupied traps.\\
The complete picture complicates if we attempt to account for the effects produced by radiation damage, which can create new defects or alter existing ones.
In extending the model to consider these situations we are strongly constrained by the available data since the afterglow emission from the crystal was not characterized before the campaign, and we performed only six irradiation steps with large changes in the proton flux. These limitations force us to introduce only a simple modification to the assumptions we made: we allow the capture probability $p_i$ to change with time during a stimulation. Assuming the change to be sufficiently small, we can expand the capture rate in a Taylor series of time, keeping only the linear term
$$ n_i(t) \approx n_{i,0}+\frac{d}{dt}n_i(t)\,t $$
We consider first the case in which radiation damage is absent or negligible. The main cause of change in the capture rates is the reduction in available empty traps due to trapping of the free charge carriers. As a consequence, neglecting the small number of decaying metastable states, we may write the approximation
$$ n_i(t) = n_i^0+k_i\,\phi\, t$$
where $k_i$ is a negative parameter. After the end of the stimulation the traps emit electrons so that at the beginning of the following stimulation the capture rate has partially recovered following the exponential decay of the metastable states.\\
Besides the mechanism just described, the change in the capture rate may be brought about by an increase in the electrons available for capture due to the reduction of trapping by another species, however this effect is quite small, since $p_i \ll 1$, and it can be observed only if the previous mechanism is negligible for the trap species under consideration.\\
In this approximation the time evolution of the number of occupied traps of the $i$-th species during the $j$-th stimulation is found by solving the system of equations
\begin{equation}
    \begin{cases}
        \frac{d}{dt}N_i(t) = n_{i,j}(t)\phi_j-\frac{N_i(t)}{\tau_i}\\
        n_{i,j}(t) = n_{i,j}^0+k_i\,\phi_j\, t
    \end{cases}
    \label{eq:model2_de}
\end{equation}
where $n_{i,j}^0$ is the value of the capture rate at the beginning of the $j$-th stimulation. The solution is
$$ N_{i,j}(t) = N_{i,j}^0\;e^{-t/\tau_i}+\tau_i\phi_j n_{i,j}^0 \left( 1-e^{-t/\tau_i} \right) + k_i\tau_i^2\phi_j^2 \left[ \frac{t}{\tau_i}-\left( 1-e^{-t/\tau_i} \right) \right] $$
where $N_{i,j}^0$ is the number of traps already occupied at the beginning of the stimulation. The first two terms on the RHS are equal to the result of the previous model, Eq. \ref{eq:model_res}, while the last term accounts for the variation in the capture rate of the trap species under consideration.\\
To write a complete formula for the evolution of the number of occupied traps we have to explicitly write the dependence of the trap rate on the previous stimulation steps in the second equation of system \ref{eq:model2_de}. By introducing the fluence $\Phi_j = \phi_j \Delta t_j$ received by the crystal in the $j$-th stimulation (defining $\Phi_0 = 0$), the time interval $\Delta t_{j-1,j}$ between the end of the $(j-1)$-th stimulation and the beginning of the $j$-th one (with $\Delta t_{0,1} = t_1$), the sum of these intervals
$$ \Delta t_{k,j} = \sum_{l=k}^{j-1} \Delta t_{l,l+1} \;\text{,}$$
and by replacing the parameter $k_i$ with $\Delta n_i/\Phi_{tot} = \Delta n_i/\sum_j \Phi_j$, to ease the interpretation of the results, we obtain
$$ n_{i,j}^0 = n_i^0 + \frac{\Delta n_i}{\Phi_{tot}} \sum_{k=0}^{j-1} \Phi_k e^{-\frac{\Delta t_{k,j}}{\tau_i}} \;\text{.}$$
At the end of the $j$-th stimulation, then, the number of newly occupied traps is given by
\begin{equation}
    \Delta N_{i,j}^{\phi} = \tau_i \phi_j \Bigg\lbrace \left[ n_i^0 + \frac{\Delta n_i}{\Phi_{tot}} \left( \sum_{k=0}^{j-1} \Phi_{k} e^{-\frac{\Delta t_{k,j}}{\tau_i}} - \tau_i\phi_j \right) \right] \left( 1-e^{-\frac{\Delta t_j}{\tau_i}} \right) + \frac{\Delta n_i}{\Phi_{tot}} \Phi_j \Bigg\rbrace
    \label{eq:DN_flux}
\end{equation}
and the time evolution of the occupied traps of the $i$-th species is still given by Eq. \ref{eq:trapag} if we replace $\Delta N_{i,j}$ with $\Delta N_{i,j}^{\phi}$.
We remark the fact that the effect becomes cumulative for trap species with mean lifetimes large with respect to the complete duration of the crystal stimulation.\\
We consider now the case in which the occupied traps can be neglected, due to a sufficiently low flux, but the radiation damage affects afterglow emission profile. The interaction of the radiation with the material making up the scintillator causes an increase in the number of defects present in the crystal, and possibly changes their densities in different ways. As a consequence, see Eq. \ref{eq:prob}, the capture probabilities of the different trap species changes: they increase in the species that feature an augmented density, and they  decrease slightly in the species whose densities are not affected by radiation damage. The phenomenon is cumulative, and at small irradiation levels the trap densities change linearly with the fluence received by the crystal. During irradiation, the capture rates are expected to change linearly with time because the radiation flux is constant, and the electron trapping is again described by the system of equations \ref{eq:model2_de}, but now the time evolution of the capture rate is replaced by
$$ n_{i,j}(t) = n_i^0 + \frac{\Delta n_i}{\Phi_{tot}} \left( \sum_{k=0}^{j-1}\Phi_k+\phi_j t \right) $$
where we used the same normalization of the rate of change in $n_i$.
The number of newly occupied traps of the i-th species at the end of the $j$-th irradiation is now
\begin{equation}
    \Delta N_{i,j}^{\Phi} = \tau_i \phi_j \Bigg\lbrace \left[ n_i^0 + \frac{\Delta n_i}{\Phi_{tot}} \left( \sum_{k=0}^{j-1} \Phi_{k} - \tau_i\phi_j \right) \right] \left( 1-e^{-\frac{\Delta t_j}{\tau_i}} \right) + \frac{\Delta n_i}{\Phi_{tot}} \Phi_j \Bigg\rbrace
    \label{eq:DN_fluence}
\end{equation}
By replacing $\Delta N_{i,j}$ with this quantities in Eq. \ref{eq:trapag} we get the time evolution of the occupied traps of the $i$-th species during the irradiation campaign, valid outside the irradiation steps.\\
Having modeled the electron capture in the traps we are now able to calculate the afterglow emission of the crystal, which is due to recombination of the charge carriers liberated in the deexcitation of the metastable states. For each trap species we choose the appropriate model, by using Eq. \ref{eq:DN_fixed}, \ref{eq:DN_flux} or \ref{eq:DN_fluence} in Eq. \ref{eq:trapag}, driven by the information contained in the data. By neglecting retrapping, the photon flux emitted by the crystal is
\begin{equation}
    \phi_q(t) = -\sum_{i=1}^{N_t} \frac{d}{dt}N_i(t) \;\text{.}
    \label{eq:photon_flux}
\end{equation}
We observe two facts. In situations like the one we are analysing, in which large steps in fluence are obtained by corresponding increases in the flux, it may be difficult to distinguish between radiation damage and a reduced availability of empty traps, because most of the effect is due to a single irradiation and both scenarios are compatible by adequately scaling the parameter $\Delta n_i$. In this case help may come from the observation of a trap species with large increase in the capture rates that is not balanced by the reductions in the capture by other species, since this is only compatible with radiation damage (also the capture rates of both unobserved trap species and recombination centers have to decrease slightly).\\
The second observation concerns the trap species with smallest mean lifetime. The models we formulated take into account only the trap species with observable mean lifetime, while we know that there certainly are traps that decay faster than the detection limit of the experiment. The emission tails of these traps influence the shape of the first portion of the measured light curve, so the parameters we get for the species with smallest mean lifetime by fitting the model to the data are offset from the real values. This fact is unavoidable, because we don't have a description of what happens before the first data point, and it hinders the interpretation of the results for this trap species. The phenomenon propagates to the other trap species, with an effect that vanishes rapidly as the mean lifetime increases.

\bibliography{mybibfile}

\begin{thebibliography}{10}
\expandafter\ifx\csname url\endcsname\relax
  \def\url#1{\texttt{#1}}\fi
\expandafter\ifx\csname urlprefix\endcsname\relax\def\urlprefix{URL }\fi
\expandafter\ifx\csname href\endcsname\relax
  \def\href#1#2{#2} \def\path#1{#1}\fi

\bibitem{Campana:2017jls}
R.~Campana, et~al., {A compact and modular X and gamma-ray detector with a CsI
  scintillator and double-readout Silicon Drift Detectors}, Proc. SPIE Int.
  Soc. Opt. Eng. 9905 (2016) 99056I.
\newblock \href {http://arxiv.org/abs/1704.06122} {\path{arXiv:1704.06122}},
  \href {https://doi.org/10.1117/12.2232597} {\path{doi:10.1117/12.2232597}}.

\bibitem{KAMADA201288}
K.~Kamada, T.~Yanagida, T.~Endo, K.~Tsutumi, Y.~Usuki, M.~Nikl, Y.~Fujimoto,
  A.~Fukabori, A.~Yoshikawa,
  \href{http://www.sciencedirect.com/science/article/pii/S0022024811010153}{2inch
  diameter single crystal growth and scintillation properties of
  ce:gd3al2ga3o12}, Journal of Crystal Growth 352~(1) (2012) 88 -- 90, the
  Proceedings of the 18th American Conference on Crystal Growth and Epitaxy.
\newblock \href
  {https://doi.org/https://doi.org/10.1016/j.jcrysgro.2011.11.085}
  {\path{doi:https://doi.org/10.1016/j.jcrysgro.2011.11.085}}.
\newline\urlprefix\url{http://www.sciencedirect.com/science/article/pii/S0022024811010153}

\bibitem{LUCCHINI2016176}
M.~Lucchini, V.~Babin, P.~Bohacek, S.~Gundacker, K.~Kamada, M.~Nikl,
  A.~Petrosyan, A.~Yoshikawa, E.~Auffray,
  \href{http://www.sciencedirect.com/science/article/pii/S0168900216001480}{Effect
  of mg2+ ions co-doping on timing performance and radiation tolerance of
  cerium doped gd3al2ga3o12 crystals}, Nuclear Instruments and Methods in
  Physics Research Section A: Accelerators, Spectrometers, Detectors and
  Associated Equipment 816 (2016) 176 -- 183.
\newblock \href {https://doi.org/https://doi.org/10.1016/j.nima.2016.02.004}
  {\path{doi:https://doi.org/10.1016/j.nima.2016.02.004}}.
\newline\urlprefix\url{http://www.sciencedirect.com/science/article/pii/S0168900216001480}

\bibitem{irenemodel}
G.~Ginet, T.~O'Brien, S.~Huston, W.~Johnston, T.~Guild, R.~Friedel,
  C.~Lindstrom, C.~Roth, P.~Whelan, R.~Quinn, D.~Madden, S.~Morley, Y.-J. Su,
  Ae9, ap9 and spm: New models for specifying the trapped energetic particle
  and space plasma environment, Space Science Reviews 179 (11 2013).
\newblock \href {https://doi.org/10.1007/s11214-013-9964y}
  {\path{doi:10.1007/s11214-013-9964y}}.

\bibitem{TPTC}
F.~{Tommasino}, M.~{Rovituso}, S.~{Fabiano}, S.~{Piffer}, C.~{Manea},
  S.~{Lorentini}, S.~{Lanzone}, Z.~{Wang}, M.~{Pasini}, W.~J. {Burger}, C.~{La
  Tessa}, E.~{Scifoni}, M.~{Schwarz}, M.~{Durante}, {Proton beam
  characterization in the experimental room of the Trento Proton Therapy
  facility}, Nuclear Instruments and Methods in Physics Research A 869 (2017)
  15--20.
\newblock \href {https://doi.org/10.1016/j.nima.2017.06.017}
  {\path{doi:10.1016/j.nima.2017.06.017}}.

\bibitem{knoll}
G.~F. Knoll, Radiation Detection and Measurement, 4th Edition, John Wiley and
  Sons, 2010.

\bibitem{2006:huntley}
D.~{Huntley}, {An explanation of the power-law decay of luminescence}, Journal
  of Physics Condensed Matter 18~(4) (2006) 1359--1365.
\newblock \href {https://doi.org/10.1088/0953-8984/18/4/020}
  {\path{doi:10.1088/0953-8984/18/4/020}}.

\bibitem{Yoneyama_2018}
M.~Yoneyama, J.~Kataoka, M.~Arimoto, T.~Masuda, M.~Yoshino, K.~Kamada,
  A.~Yoshikawa, H.~Sato, Y.~Usuki,
  \href{https://doi.org/10.1088%2F1748-0221%2F13%2F02%2Fp02023}{Evaluation of
  {GAGG}:ce scintillators for future space applications}, Journal of
  Instrumentation 13~(02) (2018) P02023--P02023.
\newblock \href {https://doi.org/10.1088/1748-0221/13/02/p02023}
  {\path{doi:10.1088/1748-0221/13/02/p02023}}.
\newline\urlprefix\url{https://doi.org/10.1088%2F1748-0221%2F13%2F02%2Fp02023}

\bibitem{sakano2014}
M.~Sakano, T.~Nakamori, S.~Gunji, J.~Katagiri, S.~Kimura, S.~Otake,
  H.~Kitamura, Estimating the radiative activation characteristics of a
  gd3al2ga3o12:ce scintillator in low earth orbit, Journal of Instrumentation 9
  (2014) P10003.
\newblock \href {https://doi.org/10.1088/1748-0221/9/10/P10003}
  {\path{doi:10.1088/1748-0221/9/10/P10003}}.

\bibitem{hamamatsu_tubes}
H.~P. K.K., Photomultiplier Tubes, Basics and Applications, 3rd Edition,
  Hamamatsu Photonics K.K., 2007.

\bibitem{gaggJ2}
C\&a gagg scintillation properties,
  \url{http://www.c-and-a.jp/assets/img/products/101180228_GAGG.pdf}, accessed:
  2020-01-29.

\bibitem{alenkov2019}
V.~Alenkov, et~al., {Irradiation studies of a multi-doped
  Gd$_3$Al$_2$Ga$_3$O$_{12}$ scintillator}, Nucl. Instrum. Meth. A 916 (2019)
  226--229.
\newblock \href {https://doi.org/10.1016/j.nima.2018.11.101}
  {\path{doi:10.1016/j.nima.2018.11.101}}.

\bibitem{campana2020}
R.~Campana, F.~Fuschino, Y.~Evangelista, G.~Dilillo, F.~Fiore,
  \href{https://doi.org/10.1117/12.2560365}{{The HERMES-TP/SP background and
  response simulations}}, in: J.-W.~A. den Herder, S.~Nikzad, K.~Nakazawa
  (Eds.), Space Telescopes and Instrumentation 2020: Ultraviolet to Gamma Ray,
  Vol. 11444, International Society for Optics and Photonics, SPIE, 2020, pp.
  817 -- 824.
\newline\urlprefix\url{https://doi.org/10.1117/12.2560365}

\bibitem{misc:esa_models}
Ecss-e-st-10-04c – space environment,
  \url{https://ecss.nl/standard/ecss-e-st-10-04c-space-environment/}, accessed:
  2020-09-25.

\bibitem{2014:campanasax}
R.~{Campana}, M.~{Orlandini}, E.~{Del Monte}, M.~{Feroci}, F.~{Frontera}, {The
  radiation environment in a low earth orbit:the case of BeppoSAX},
  Experimental Astronomy 37~(3) (2014) 599--613.
\newblock \href {http://arxiv.org/abs/1405.0360} {\path{arXiv:1405.0360}},
  \href {https://doi.org/10.1007/s10686-014-9394-1}
  {\path{doi:10.1007/s10686-014-9394-1}}.

\bibitem{ripa2020}
J.~{{\v{R}}{\'\i}pa}, G.~{Dilillo}, R.~{Campana}, G.~{Galg{\'o}czi}, {A
  comparison of trapped particle models in low Earth orbit}, in: Society of
  Photo-Optical Instrumentation Engineers (SPIE) Conference Series, Vol. 11444
  of Society of Photo-Optical Instrumentation Engineers (SPIE) Conference
  Series, 2020, p. 114443P.
\newblock \href {http://arxiv.org/abs/2101.03090} {\path{arXiv:2101.03090}},
  \href {https://doi.org/10.1117/12.2561011} {\path{doi:10.1117/12.2561011}}.

\end{thebibliography}

\end{document}